\documentclass[aps,prb,superscriptaddress,tightenlines,nofootinbib,floatfix,longbibliography,notitlepage]{revtex4-2}

\usepackage{graphicx}
\usepackage[T1]{fontenc}
\usepackage[utf8]{inputenc}
\usepackage[UKenglish]{babel}
\usepackage{placeins}
\usepackage[shortlabels]{enumitem}
\usepackage[binary-units=true]{siunitx}
\usepackage{color}
\usepackage[dvipsnames]{xcolor}
\usepackage{amsmath}
\usepackage{amssymb}
\usepackage{amsthm}
\usepackage{mathtools}
\usepackage{bbold}
\usepackage{dsfont}
\usepackage{braket}
\usepackage[ruled,lined]{algorithm2e}
\usepackage{nicefrac}
\usepackage[pdftex,
bookmarks,
urlcolor=blue,
colorlinks]{hyperref}
\usepackage{cleveref}

\graphicspath{{figures/},{data/}}

\crefformat{footnote}{#2\footnotemark[#1]#3}

\DeclareMathOperator{\diag}{diag}

\DeclareMathOperator{\ord}{\mathcal{O}}

\newcommand{\matr}[2]{\left(\begin{array}{#1}#2\end{array}\right)}

\newcommand{\eto}[1]{\ensuremath{e^{#1}}}

\newcommand{\ordnung}[1]{\ensuremath{\ord\left(#1\right)}}

\newcommand{\br}[1]{\ensuremath{\left(#1\right)}}
\newcommand{\brr}[1]{\ensuremath{\left[#1\right]}}
\DeclarePairedDelimiter\abs{|}{|}

\colorlet{myorange}{orange}
\colorlet{mypurple}{purple}
\definecolor{myblue}{HTML}{1E88E5}
\definecolor{mygreen}{HTML}{004D40}

\theoremstyle{definition}

\theoremstyle{remark}

\usepackage{simpler-wick}

\usepackage{color}

\newcommand{\CombineUDT}{\textsc{CombineUDT}}
\SetKwFunction{FQR}{QR}
\SetKwFunction{Fdiag}{diag}
\SetKwFunction{FCombineUDT}{\CombineUDT}
\SetKwInOut{Input}{input}
\SetKwInOut{Params}{parameters}
\SetKwInOut{Output}{output}
\SetKw{KwReturn}{return}

\providecommand{\orgname}[1]{#1}
\providecommand{\orgaddress}[1]{#1}
\providecommand{\postcode}[1]{#1}
\providecommand{\city}[1]{#1}
\providecommand{\country}[1]{#1}

\newcommand{\bonn}{
    \orgname{%
        Helmholtz-Institut f\"{u}r Strahlen- und Kernphysik, %
        Rheinische Friedrich-Wilhelms-Universit\"{a}t 
    }%
    \orgaddress{%
        \postcode{53115} \city{Bonn}, \country{Germany}
    }%
}

\newcommand{\ias}{
    \orgname{%
        Institute for Advanced Simulation 4, %
	    Forschungszentrum J\"{u}lich,
    }%
    \orgaddress{%
        \postcode{52428} \city{J\"{u}lich}, \country{Germany}
    }%
}

\makeatletter%
\makeatletter%
\begin{document}

	\title{Stable Determinant Monte Carlo Simulations at Low Temperatures}

\author{Thomas Luu}
	\email{t.luu@fz-juelich.de}
	\affiliation{\ias}
	\affiliation{\bonn}
\author{Johann Ostmeyer}
	\email{ostmeyer@hiskp.uni-bonn.de}
	\affiliation{\bonn}
\author{Petar Sinilkov}
	\email{p.sinilkov@fz-juelich.de}
	\affiliation{\ias}
\author{Finn L. Temmen}
	\email{f.temmen@fz-juelich.de}
	\affiliation{\ias}

 	\date{\today}
	
	\begin{abstract}
	At low temperatures $T$ where $\nicefrac{1}{T}=\beta\gg1$ the na\"ive implementation of determinant quantum Monte Carlo (DQMC) methods suffers from loss of precision and numerical instabilities when evaluating the fermion determinant.
	In double precision the simulations of systems with a band width $\Delta$ deteriorate around $\beta \Delta \approx 70$.
	This instability propagates into the calculation of observables that rely on the evaluation of the inverse of the fermion matrix, or the Green's function.  
	For DQMC methods that rely on the Hamiltonian Monte Carlo (HMC) algorithm, an additional complication comes from evaluating the force terms required for integrating Hamilton's equations of motion, since here loss of precision and numerical instabilities are also prevalent.
	We show how to address all these issues using various choices of matrix decompositions, allowing us to simulate at $\beta\gtrsim 100$ or $\beta\Delta\gtrsim 600$, which corresponds to room temperature for graphene structures.  
	Furthermore, our implementation has numerical costs that scale similarly to the na\"ive implementation,  namely as $\mathcal{O}(N_x^3N_t)$, where $N_x$ ($N_t$) is the number of spatial (temporal) sites.
	\end{abstract}

	\maketitle
	\newpage
	\tableofcontents
	\newpage

	\unitlength = 1em
	\section{Introduction}
	Numerical simulations play a central role in advancing our understanding of strongly correlated electron systems and are an essential tool for studying chemical molecules and materials.
	In condensed matter physics, such systems are commonly described by quantum many-body models of interacting electrons, including the Hubbard model, the Pariser-Parr-Pople model, and related lattice Hamiltonians. 
	Among the most successful approaches for simulating these models are quantum Monte Carlo (QMC) methods, which reformulate expectation values as a stochastic sampling problem over auxiliary fields and analytically integrate out the fermionic degrees of freedom, resulting in a fermion determinant.
	For this reason, QMC methods are also commonly referred to as determinant quantum Monte Carlo (DQMC) methods.
	Several algorithms exist within this framework, most prominently the Blankenbecler-Scalapino-Sugar (BSS)~\cite{BSS:1981} algorithm and the Hybrid/Hamiltonian Monte Carlo (HMC) method \cite{HMC} enhanced with radial updates \cite{original_radial_update,Ostmeyer:2024gnh, Temmen_ergodicity}; these methods have been successfully applied to a wide range of strongly correlated systems, from graphene and electron-phonon models to Mott insulators and superconductors \cite{Brower:2011av,ConfPhaseTransitionGraphene,KRIEG201915,HofmannTBG,HMC_MonolayerGraphene,Revisiting_HQMC_for_Hubbard, CohenSteadElectronPhononHMC, AFM_SSHH, CDW_ElectronPhonon,OstmeyerMottInsulator,DQMC_OSMT,highTcSuperconductors,HubbardBilayer, HMCExtendedHubbardModel,QuantumCriticalityDiracSystems,du2026finitetemperaturequantummontecarlo}.

	However, independently of the specific algorithm, accurately simulating these systems at physically relevant temperatures remains a major computational challenge.
	While the fermionic sign problem is well known to cause exponentially growing computational costs with decreasing temperature, DQMC simulations face a more fundamental numerical obstacle: the finite precision inherent to floating point arithmetic used in the underlying matrix operations.
	Specifically, the core computational task involves repeated multiplication of numbers with vastly different scales, with this scale separation growing exponentially as temperature decreases.
	Rounding errors from these operations then accumulate rapidly, and without proper stabilization they can silently distort physical observables or, in more severe cases, cause complete algorithmic breakdown.

	Therefore, stabilization methods are essential for any reliable DQMC simulation at low temperatures, and a well-established approach based on scale separation has emerged in the literature \cite{AlgorithmForTheSimulationOfManyElectronSystems,Bauer:2020stable,Kreit:2026eng, AssaadEvertz2008, Loh:2005, PhysRevB.99.045108}.
	The aim of the present paper is to extend these ideas to the stable and efficient computation of Green's functions, which are needed for computing observables in general DQMC setups and, more specifically, for the closely related force term evaluation in HMC.
	
	To this end, we revisit the stabilization schemes presented in Ref.~\cite{Bauer:2020stable}, which serve as a starting point for our improvements, and briefly comment on efficient parallelization strategies (see sec.~\ref{sec:stable-logdet}).
	We then extend this framework by first demonstrating that a na\"ive application of the stabilization scheme is insufficient for Green's function evaluation, using force term computations as our primary example (see sec.~\ref{sec:naive-grad}).
	Subsequently, we present a significantly more stable approach that addresses the encountered limitations (see sec.~\ref{sec:stable-grad}).
	While we focus on force terms for concreteness, we stress that these techniques apply equally to the measurement of other observables like correlation functions in any DQMC simulation.
	In fact, we continue by demonstrating how the same intermediate quantities used for stable force computations can be efficiently exploited to compute the full Green's function (see sec.~\ref{sec:stable-prop}).
	These improvements benefit all DQMC calculations, but are essential for HMC simulations, where unstable force evaluation causes algorithmic breakdown at larger inverse temperatures.
	Our results provide a comprehensive framework for stabilizing both measurements and HMC simulations, significantly extending the accessible parameter space.
	The framework is agnostic to the spatial dimensionality and geometry of the underlying system, enabling its application to a broad range of materials and lattice models, including room-temperature simulations of graphene and warm molecules such as Perylene or Corannulene.
	
	These results have been collected into a handbook~\cite{Ostmeyer:2026sea} together with a number of other algorithms for the fermion determinant.
	We recommend to check it out before implementing any specific algorithm for the simulation of fermionic systems.
	Example implementations of DQMC simulations can be found in Refs.~\cite{ALF/SciPostPhysCodeb.1-v2.4,10.21468/SciPostPhysCodeb.29,NSL}.
	
	\section{Formalism}	
	
	In DQMC simulations of condensed matter systems with continuous variables\footnote{The matrix operations discussed in this work can be stabilised in much the same way for discrete fields like those in Hirsch's formulation~\cite{Hirsch:1983zza}.}
	the partition function is recast as an integral over an auxiliary field $\phi_{x,t}$ at each space-time point~\cite{Brower:2011av},
	\begin{equation}\label{eqn:partition}
	Z=\int\mathcal{D}[\phi]\left(\prod_{f=1,2}\det M^{(f)}[\phi]\right)e^{-S[\phi]}=
	\int\mathcal{D}[\phi]e^{-S[\phi]+\sum_{f=1,2}\log\det M^{(f)}[\phi]}\equiv
	\int\mathcal{D}[\phi]e^{-S_\text{eff}[\phi]}\ ,
	\end{equation}
	where the index $f$ represents the different fermion species (e.g.\@ spin $\uparrow$ and $\downarrow$ fermions in the \emph{spin} basis, or `particle' and `hole' fermions in the \emph{charge}, or \emph{particle-hole} basis, etc.), 
	\begin{displaymath}
	\mathcal{D}[\phi]\equiv\prod_{x,t}d\phi_{x,t}\ ,
	\end{displaymath}
	and the `action' $S[\phi]$ is a functional of the field $\phi$.
	For example, the Hubbard model at half-filling described by the Hamiltonian\footnote{The $\mp$ sign in~\eqref{eqn:Hstart} demarcates the `spin basis' and the `charge basis', respectively. See Ref.~\cite{Wynen:2018ryx} for details.}
	\begin{equation}
	\label{eqn:Hstart}
	H = -\sum_{x,y,f=f_1,f_2}c^\dag_{x,f}\kappa_{x,y}c^{}_{y,f}\mp\frac{U}{2}\sum_x\left(c^\dag_{x,f_1}c^{}_{x,f_1}-c^\dag_{x,f_2}c^{}_{x,f_2}\right)^2\ ,
	\end{equation}
	where $\kappa$ is the connectivity matrix and $U$ is the onsite interaction, has the action
	\begin{equation}\label{eqn:Hubbard action}
	S[\phi]=\frac{1}{2}\sum_{x,t}\frac{\phi^2_{x,t}}{U\delta}\ ,
	\end{equation}
	where $\delta =\beta/N_t$ with $\beta$ being the inverse temperature and $N_t$ the number of timeslices.
	For our work we will assume our fermions are dynamically described by a real and symmetric $\kappa$ matrix.
	We emphasize, however, that the formalism makes no assumptions about the spatial dimensionality or geometry of the system.
	The lattice enters solely through $\kappa$, making the framework applicable to arbitrary spatial dimensions and lattice geometries.

	The fermion matrix $M$ in~\eqref{eqn:partition} plays a central role in DQMC.
	As its derivation is found commonly in various references (e.g.~\cite{Luu:2015gpl,Wynen:2018ryx}), we only state its general block-matrix form here:
	\begin{align}\label{eqn:fermion matrix}
	M &= \begin{pmatrix}
		\mathbb{1} & 0 & \cdots & 0 & M_{N_t-1} \\
		-M_0 & \mathbb{1} & 0 & \cdots & 0 \\
		0 & -M_1 & \mathbb{1} & \ddots & \vdots \\
		\vdots & \ddots & \ddots & \ddots & 0 \\
		0 & \cdots & 0 & -M_{N_t-2} & \mathbb{1} 
		\end{pmatrix}\,.
	\end{align}
	Here we have
	\begin{equation}\label{eqn:Ft}
	M_t\equiv \exp\left(\delta \kappa\right)\cdot {\rm diag}\left(e^{i\phi_{0,t}},e^{i\phi_{1,t}},\ldots,e^{i\phi_{x,t}},\ldots,e^{i\phi_{N_x-1,t}}\right)\ .
	\end{equation}
	This expression is valid in the \emph{charge} basis.
	To obtain the equivalent expression in the \emph{spin} basis, it suffices to let $\phi_{x,t}\to-i\phi_{x,t}\ \forall\ x,t$ in~\eqref{eqn:Ft}.
	Each term in~\eqref{eqn:fermion matrix} represents an $N_x\times N_x$ matrix, and the location and different sign associated with $M_{N_t-1}$ are due to anti-periodic temporal boundary conditions of the fermions.
	We note that the non-interacting limit $U\to0$ results in $\phi_{x,t}=0\ \forall\ x,t$.
	
	\subsection{Stable calculation of $\log\det(M)$}
	\label{sec:stable-logdet}
	As its name suggests, the calculation of the (logarithmic) determinant of the fermion matrix $M$ is required for DQMC.  
	By application of Sylvester's determinant identity~\cite{AKRITAS1996585} and following the steps detailed e.g.\@ in Ref.~\cite{Wynen:2018ryx}, it is easy to show that 
	\begin{equation}\label{eqn:sylvester}
	\det(M) = \det\left(\mathbb{1}+M_{0} M_{1}\ldots M_{N_t-1}\right)\ .
	\end{equation}	
	Before describing the general procedure for evaluating this determinant, it is worthwhile to consider this expression in the non-interacting limit where $\phi_{x,t}=0\ \forall\ x,t$.
	The product $\prod_tM_{t}$ simplifies to $\exp\left(\beta\kappa\right)$ and the determinant can be analytically determined by going to the diagonal basis of $\kappa$, giving
	\begin{displaymath}
	\left.\det(M)\right\vert_{U=0}=\prod_i\left(1+e^{\beta\lambda_i}\right)\ ,
	\end{displaymath}
	where $\lambda_i$ is the $i^\text{th}$ eigenvalue of $\kappa$.
	Note that for an adjacency matrix $\kappa$ with number of nearest neighbors\footnote{The 2D honeycomb lattice has $K=3$, while the 2D square lattice has $K=4$.} $K$ the eigenvalues of $\kappa$ are constrained to $-K\le \lambda_i\le K$.  
	This means that the condition number of the matrix $\exp\left(\beta\kappa\right)$, which is the ratio of its largest to smallest eigenvalues, is $e^{2\beta K}$.
	Thus the condition number grows exponentially in $\beta$.
	This fundamentally is the reason for numerical instabilities in any na\"ive calculation of the determinant.
	
	The interacting case affords no simple expression.  As is well documented in Ref.~\cite{Bauer:2020stable}, a na\"ive construction of $M_0\ldots M_{N_t-1}$ by matrix-matrix multiplication suffers from numerical instabilities when $K\beta\gtrsim 35$ in a non-interacting system.  
	Fortunately Ref.~\cite{Bauer:2020stable} provides a clever algorithm for stabilizing the determinant based on recursive matrix decompositions, such as \texttt{QR} and \texttt{SVD}, of the products involving $M_t$.
	The method of recursive decompositions will serve as a central tool throughout this work and we briefly review it in the following paragraph.
	We also include the \texttt{SVD} in our discussions because it is often more intuitive.
	However, \texttt{QR}-based decompositions are both faster and numerically more stable.
	Therefore, in practical implementations \texttt{QR} should always be used.

	Our starting point is either an initial \texttt{QR} or \texttt{SVD} decomposition of each $M_t$, which we express as $M_t =  U_t D_t T_t$.  
	If \texttt{SVD} is employed, then $U_t$ and $T_t$  are unitary matrices and $D_t$ is diagonal.
	If instead we use $\texttt{QR}$, then $U_t$ is unitary, $D_t$ is diagonal, and $T_t$ is triangular.
	For the typical \texttt{QR} decomposition $M=Q R$ of a square matrix $M$, where $Q$ is unitary and $R$ is triangular, the matrices $U D T $ are obtained via $U=Q$, $D={\rm diag}(R)$, and $T=D^{-1} R$.
	The decomposition can be done quickly and efficiently if we first precompute and store the decomposition $\exp\left(\delta\kappa\right)=U_\kappa D_\kappa T_\kappa$.
	Then from~\eqref{eqn:Ft} it follows that $U_t=U_\kappa$, $D_t=D_\kappa$, and $T_t=T_\kappa\cdot{\rm diag}\left(e^{i\phi_{0,t}},e^{i\phi_{1,t}},\ldots,e^{i\phi_{N_x-1,t}}\right)$, 
	the last of which can be quickly evaluated for varying $\phi_{x,t}$.
	We then have
	\begin{equation}\label{eqn:decomp}
	M_{0} M_{1}\ldots M_{N_t-1}=U_{0} \underbrace{D_{0} ((T_{0} U_{1}) D_{1})}_{=U D T} T_{1}\ldots U_{N_t-1} D_{N_t-1} T_{N_t-1}\ .
	\end{equation}
	We then recursively perform the same decomposition for every term of the form $D_{t'} T_{t'} U_t D_t$ (one of which is shown in~\eqref{eqn:decomp}) and recombine subsequent $U$ and $T$ matrices until we arrive at
	\begin{equation}\label{eqn:decomp2}
	M_{0} M_{1}\ldots M_{N_t-1}=U_{0} U D T T_{N_t-1}=\tilde U D \tilde T\  .
	\end{equation}
	A pseudocode implementation of the combination of two matrices via their $UDT$ decompositions is given in algorithm~\ref{alg:combined_udt}.
	Pseudocode notation and conventions used throughout this work are detailed in app.~\ref{app:pseudocode}.
		
	\begin{algorithm*}
		\caption{\CombineUDT --- Combination of two UDT decomposed matrices obtained from \textsc{QR} decomposition as introduced in Ref.~\cite{Bauer:2020stable}.}\label{alg:combined_udt}
		\Input{
			$U_0,\,D_0,\,T_0$: UDT factors of $M_0$ ($U_0$ unitary, $D_0$ diagonal, $T_0$ upper triangular) \\
			\ $U_1,\,D_1,\,T_1$: UDT factors of $M_1$
		}
		\Output{$U_{\mathrm{out}},\,D_{\mathrm{out}},\,T_{\mathrm{out}}$: UDT factors of $M_0M_1$}
		\BlankLine
		$(Q,R) \gets \FQR\bigl( D_0 ((T_0 U_1) D_1) \bigr)$\;
		$U_{\mathrm{out}} \gets U_0 Q$\;		
		$D_{\mathrm{out}} \gets \Fdiag(R)$\;
		$T_{\mathrm{out}} \gets D_{\mathrm{out}}^{-1} R$\;
		$T_{\mathrm{out}} \gets T_{\mathrm{out}} T_1$\;
		\KwReturn $U_{\mathrm{out}},\,D_{\mathrm{out}},\,T_{\mathrm{out}}$\;
	\end{algorithm*}

	In our setting, numerical stability is governed primarily by the relative scaling of the diagonal elements of the matrices, rather than by the precise order in which the matrix multiplications are performed. This observation allows the multiplications to be reordered and parallelized without compromising numerical robustness.
	
	While a straightforward sequential scan computes the required result correctly, it does not fully exploit available parallel resources. More work-efficient parallel alternatives, such as tree-based scans (e.g.\@ the Blelloch scan \cite{Blelloch1990,BlellochScan1990}) or equivalent divide-and-conquer formulations, reduce the critical path to $\ordnung{\log{n}}$ while preserving linear total work, and are well suited to both multicore CPUs and GPUs.

\begin{figure}[t!]
	\includegraphics[width=.5\textwidth]{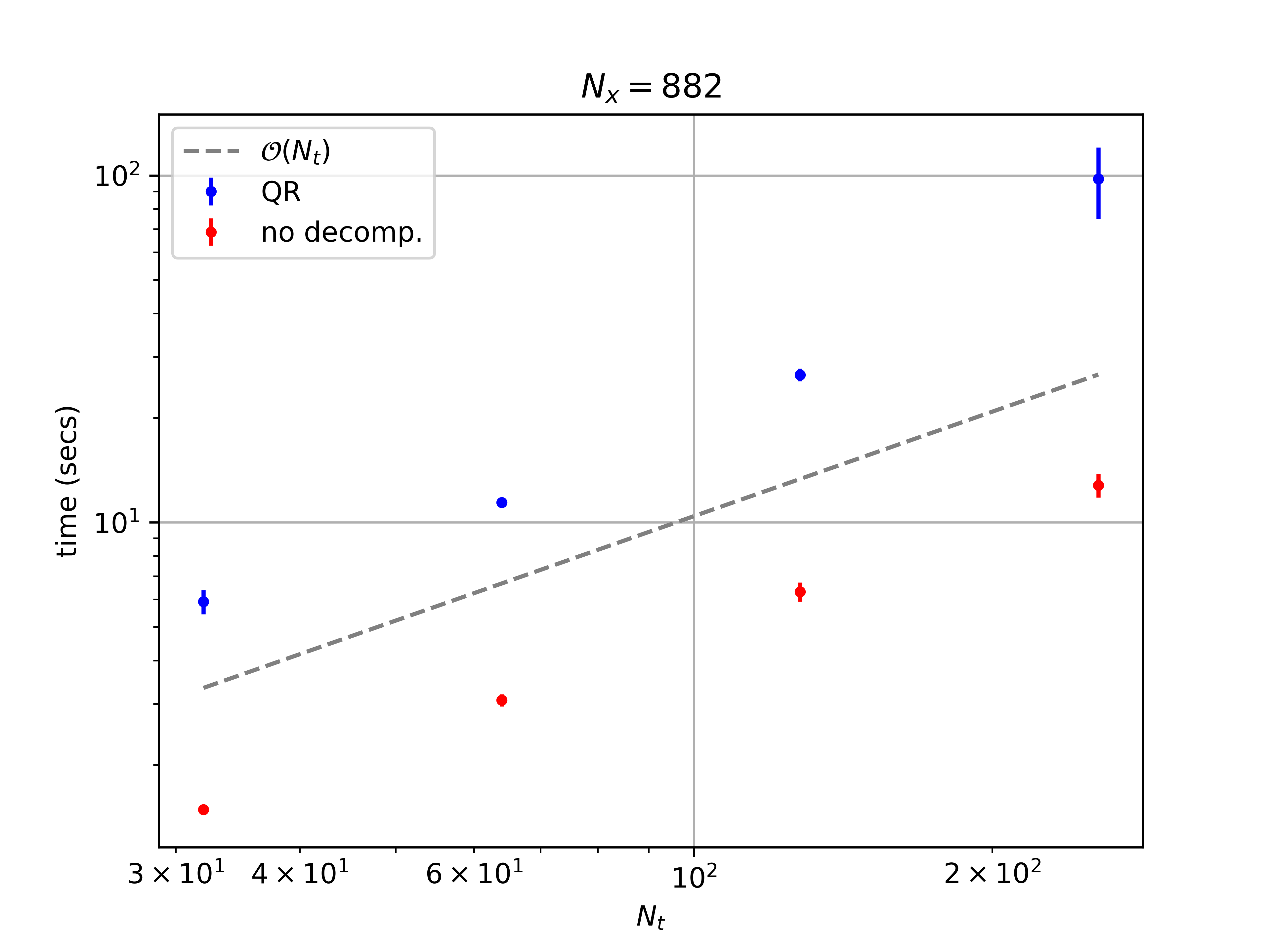}\includegraphics[width=.5\textwidth]{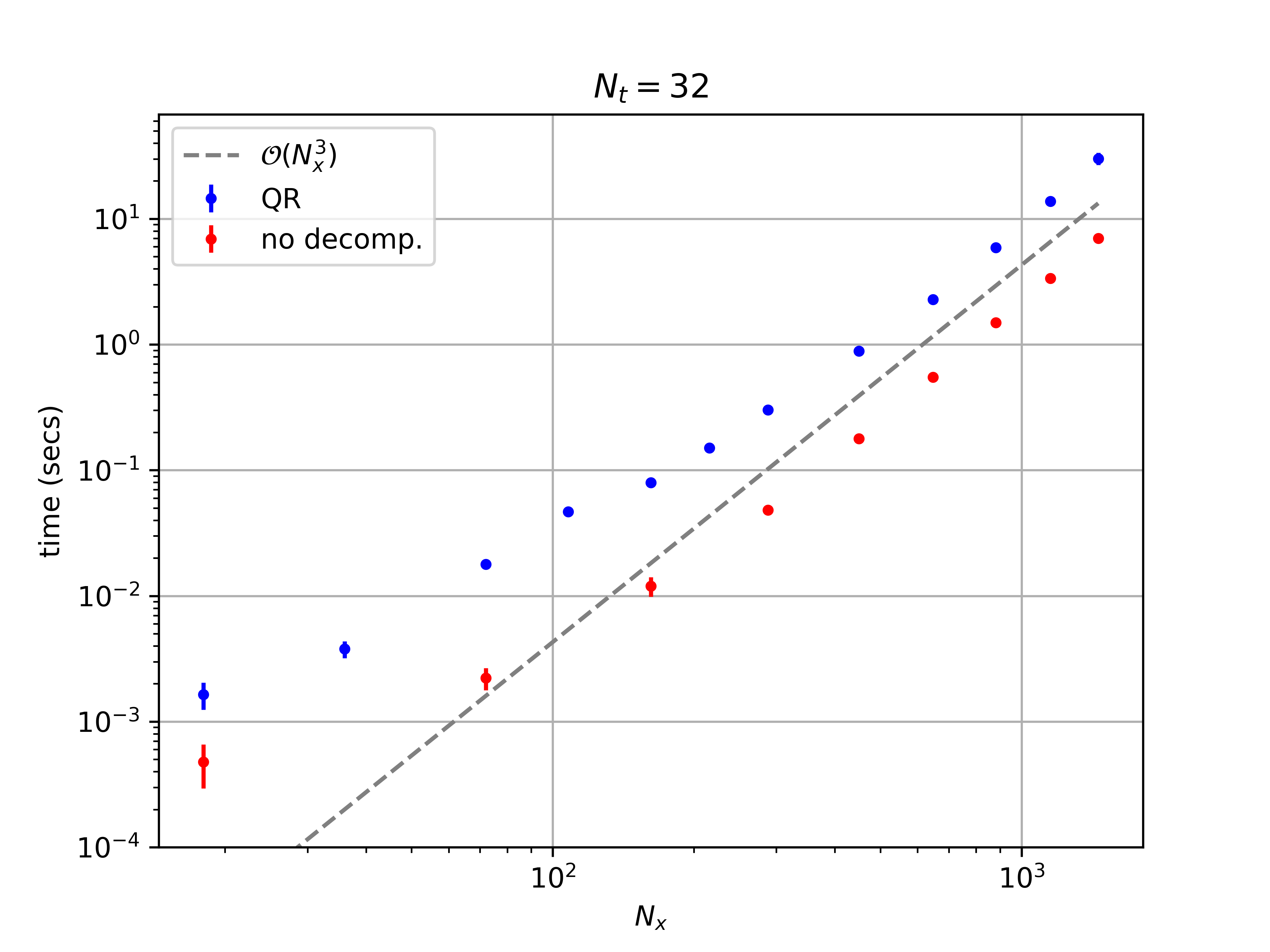}
	\caption{The scaling behavior for both decompositions (``QR'') and naïve implementation (``no decomp.'') of the evaluation of $\log\det M$.  The left panel shows the expected $\mathcal{O}(N_t)$ scaling at fixed system size $N_x$ $(=882)$, while the right panel shows the expected asymptotic $\mathcal{O}(N_x^3)$ scaling at fixed $N_t$ $(=32)$.  The gray dashed lines are to guide the eye.  Our decompositions are roughly a factor $5$ slower compared to the naïve implementation.}
	\label{fig:scaling_comparison}
	\end{figure}

	Any dense $N_x\times N_x$ matrix manipulation (matrix-matrix multiplication, \texttt{QR} or \texttt{SVD} decomposition, inversion etc.) comes at a computational complexity of order $\ordnung{N_x^3}$.
	Therefore, even the most na\"ive calculation of the fermion determinant $\det(M)$ via equation~\eqref{eqn:sylvester} has a total cost of order $\ordnung{N_t N_x^3}$.
	This is the very same scaling that we obtain for our maximally stable algorithm.  
	In fig.~\ref{fig:scaling_comparison} we show the scaling behavior of decompositions when calculating $\log\det M$ in comparison with the naïve implementation (no decompositions). 
	Here we see the expected $\mathcal{O}(N_t)$ scaling when we fix the system size $N_x$ (left panel), and conversely the expected asymptotic $\mathcal{O}(N_x^3)$ scaling with fixed $N_t$ (right panel).
	Our decompositions are roughly a factor of five slower than the naïve implementation.
	
	The stability enhancement due to these decompositions is well documented in Ref.~\cite{Bauer:2020stable}, but it is worthwhile to provide a cursory explanation here, as we will use similar arguments in the sections to come.  The origin of the enhanced stability comes from the separation of scales introduced through the $UDT$-splitting. %
	To see why, let us assume the simplest non-trivial case of a two-dimensional Hamiltonian
	\begin{align}
		H &= \br{v_1,v_2} \matr{cc}{E_1 & 0\\ 0 & E_2} \matr{c}{v_1^\dagger\\v_2^\dagger}
	\end{align}
	with the eigen-energies $E_{1,2}$ and -vectors $v_{1,2}$.
	Without loss of generality we choose $E_1 \ge E_2$.
	Then the matrix exponential related to the product $\prod_t M_t$ at a given inverse temperature $\beta$ becomes
	\begin{align}
		\eto{-\beta H} &= \br{v_1,v_2} \matr{cc}{\eto{-\beta E_1} & 0\\ 0 & \eto{-\beta E_2}} \matr{c}{v_1^\dagger\\v_2^\dagger}\\
		&= \eto{-\beta E_1} v_1 v_1^\dagger + \eto{-\beta E_2} v_2 v_2^\dagger\label{eq:2D-H-exponential}\\
		&= \eto{-\beta E_2} \br{\eto{-\beta \br{E_1-E_2}} v_1 v_1^\dagger + v_2 v_2^\dagger}\,.
	\end{align}
	Since the eigenvectors contribute only factors of order unity, the relative scale within the sum is fully determined by $\eto{-\beta \br{E_1-E_2}}$.
	Explicitly performing the summation in finite precision arithmetic therefore comes at a precision loss of this scale.
	In particular, with the machine precision $\epsilon$, all information about the $1^\text{st}$ state is entirely lost if $\eto{-\beta \br{E_1-E_2}}<\epsilon$.
	Our $UDT$-decomposition, on the other hand, keeps track of all the eigenmodes independently at their respective scales, much like the separation in~\eqref{eq:2D-H-exponential}.
	
	The argument of the determinant requires the addition of the identity matrix to this product, and this may result in precision loss and instability if done without care, even with the decompositions performed above.
	To address this, we do one final decomposition,
	\begin{equation}\label{eqn:decomp3}
	\det\left(\mathbb{1}+M_{0}\ldots M_{N_t-1}\right)=\det\left(\tilde U\tilde U^\dagger+\tilde U D \tilde T\right)=\det(\tilde U(\underbrace{\mathbb{1}+D\tilde T\tilde U}_{u d t})\tilde U^\dagger )=\det(\tilde U u d t \tilde U^\dagger).%
	\end{equation}
	When using \texttt{QR}, we can identify $D\tilde T\tilde U$ with the first step in the QR algorithm, i.e.\@ replacing $A_0=QR$ by $A_1=RQ$ which is close to diagonal.
	Both cases result in a diagonally dominant matrix whose decomposition is extremely stable.
	The logarithmic determinant is thus\footnote{In practice, when using \texttt{QR}, it is more straightforward to work directly with $Q$ and $R$ rather than introducing the intermediate factorization $QR=udt$, since $\log\det(u)+\sum_i \left(\log(d_i)+\log(t_{ii})\right) = \log\det(u)+\sum_i \log(d_i) = \log\det(Q)+\sum_i \log(R_{ii})$, where we used that $\sum_i \log(t_{ii}) = 0$ since $t_{ii} = d_i^{-1} d_i$.}
	\begin{equation}\label{eq:logdetM}
	\log\det(M)=
	\begin{cases}
	\log\det(u)+\sum_i \log(d_i) & {\rm \texttt{QR}}\ ,\\
	\log\det(u)+\sum_i \log(d_i)+\log\det(t)& {\rm \texttt{SVD}}\ .
	\end{cases}
	\end{equation}
	Here we have used the fact that the determinant of a triangular matrix is simply the product of its diagonal elements.
	Note also that since $u$ is unitary,  the determinant of this matrix has modulus $\abs*{\det(u)}=1$.
	Pseudocode for the computation of eq.~\eqref{eq:logdetM} is given in algorithm~\ref{alg:logdetM}.
	\begin{algorithm*}
		\caption{\texttt{LogDetM} --- Stable computation of the logarithm of the fermion determinant \eqref{eqn:sylvester}.}\label{alg:logdetM}
		\Params{number of time slices $N_t$}
		\Input{
			$\{U_i,\,D_i,\,T_i\}_{i=0}^{N_t-1}$: UDT factors of the matrices $\{M_i\}_{i=0}^{N_t-1}$ ($U_i$ unitary, $D_i$ diagonal, $T_i$ upper triangular) \\
			$\mathbb{1}$: Identity matrix
		}
		\Output{$\log \det M$: Logarithm of fermion determinant}
		\BlankLine
		\For{$i\gets 1$ \KwTo $N_t-1$}{
			$U_0, D_0, T_0 \gets \FCombineUDT\left( U_0, D_0, T_0, U_i, D_i, T_i \right)$\;
		}
		$(Q,R) \gets \FQR\bigl( \mathbb{1} + D_0T_0U_0 \bigr)$\;
		$\log \det M \gets \sum_i \log R_{ii} + \log\det(Q)$\;
		\tcc{Notice that $\sum_i \log R_{ii} = \sum_i \log(D_{ii})$ for $D = \Fdiag(R)$}
	\end{algorithm*}
	
	As opposed to Ref.~\cite{Bauer:2020stable}, we do not require additional stabilisation using the so-called Loh splitting~\cite{Loh:2005}.
	This makes our algorithm more transparent and completely scale-agnostic.
	The main reason for the enhanced stability of our algorithm is detailed in the next section where we show how to treat the fermionic forces (or ``time-displaced Green’s functions'').
	We will quantify these statements and provide a comparison of the methods in~\cite{Bauer:2020stable} with ours in the Sec.~\ref{sec:stable-prop}.
	We note that a na\"ive implementation of this algorithm comes at a significant runtime overhead of a factor $\ordnung{N_t}$.
	However, as we will show, calculating and storing all pre- and suffix terms (see eq.~\eqref{eqn:prefix suffix}) in advance completely negates this overhead.

	\subsection{Na\"ive calculation of $\partial_{\phi_{x,t}}\log\det(M)$}
	\label{sec:naive-grad}
	In the following we demonstrate that a na\"ive implementation of the stabilization introduced above is insufficient to stabilize the force terms and requires additional modifications.
	In DQMC simulations that rely on the HMC algorithm to generate the Markov chain, the integration of Hamilton's Equations of Motion (EoMs) is required to propose a new state.  
	To keep the presentation concise we give a short description of this algorithm in app.~\ref{app:HMC} and only describe the parts relevant to stability here.
	An essential component of evolving these EoMs is the calculation of the force term per spatial site $x$ and timeslice $t$.
	We consider the gradient of the fermion determinant
	\begin{equation}
		\dot\pi_x(t)\equiv\partial_{\phi_{x,t}}\log\det(M)\ ,
	\end{equation}
	which can be evaluated using Jacobi's formula, such that
	\begin{equation}
		\dot\pi_x(t)= \mathrm{tr}
			\left(
				\left(\mathbb{1}+M_{0} \ldots M_{N_t-1}\right)^{-1} \partial_{\phi_{x,t}} M_{0} \ldots M_{N_t-1} 
			\right)
			\ .
	\end{equation}
	At this point, we adopt the equivalent definition of eq.~\eqref{eqn:Ft},
	\begin{equation}\label{eqn:Ft_swap}
		M_t\equiv {\rm diag}\left(e^{i\phi_{0,t}},e^{i\phi_{1,t}},\ldots,e^{i\phi_{x,t}},\ldots,e^{i\phi_{N_x-1,t}}\right)\cdot \exp\left(\delta \kappa\right)\ ,
	\end{equation}
	which leaves the determinant invariant, as $\det(\mathbb{1} + AB) = \det(\mathbb{1} + BA)$.
	Defining a term based on matrix products very similar to~\eqref{eqn:sylvester},
	\begin{equation}\label{eqn:N}
	\mathcal{N}\equiv \left(\mathbb{1}+M^{-1}_{N_t-1} \ldots M^{-1}_{0}\right)^{-1}\ ,
	\end{equation}
	we obtain a simple expression for the force terms,
	\begin{equation}
	\begin{split}
	\dot\pi_x(0)&=[\mathcal{N}]_{xx}\\
	\dot\pi_x(1)&=[M^{-1}_0 \mathcal{N} M^{}_0]_{xx}\\
	&\vdots \\
	\dot\pi_x(t)&=[M^{-1}_{t-1}\ldots M^{-1}_0 \mathcal{N} M^{}_0\ldots M^{}_{t-1}]_{xx}\ .
	\end{split}\label{eq:forces_recursive}
	\end{equation}
	For brevity, in this and subsequent expressions we have omitted the constant (imaginary) prefactor arising from the derivative of matrix elements in~\eqref{eqn:Ft}.
	If we treat $\dot\pi(t)$ as an $N_x\times N_x$ matrix, we can obtain a nice recursive formula for the forces,
	\begin{equation}\label{eqn:recursive}
	\dot\pi(t)=M^{-1}_{t-1} \dot\pi(t-1) M^{}_{t-1}\quad\quad; \quad\quad\dot\pi(0)=\mathcal{N}\ .
	\end{equation}
	The force at spatial site $x$ and timeslice $t$ is then given by the diagonal matrix element of $\dot\pi(t)$.
	
	These expressions hint at a fast and efficient procedure for calculating the forces. 
	First off, $\mathcal{N}$ follows analogously to~\eqref{eqn:decomp3}, but from the inverse time slices.
	Decomposing $\mathbb{1}+M^{-1}_{N_t-1} \ldots M^{-1}_{0}=\tilde U u d t \tilde U^\dagger$ gives
	\begin{equation}\label{eqn:decomp N}
	\mathcal{N}=\tilde Ut^{-1} d^{-1} (\tilde Uu)^\dagger\ , 
	\end{equation}
	and since $\det M = \det\left( \mathbb{1}+M^{-1}_{N_t-1} \ldots M^{-1}_{0} \right) \prod_t \det M_t$, we can obtain the fermion determinant as well.
	In the case of \texttt{QR}, since $t$ is triangular, its inverse can be obtained quickly via back substitution. 
	We have verified that this calculation is numerically very stable.
	Assuming we have already precomputed and stored the initial decompositions of $M_t=U_t D_t T_t$, then it is straightforward to calculate the force terms recursively  as given in~\eqref{eqn:recursive},
	\begin{equation}\label{eqn:recursive2}
	\begin{split}
	\dot\pi(0)&=\underbrace{\tilde Ut^{-1} d^{-1} (\tilde U u)^\dagger}_{\mathcal{N}}\\
	 \dot\pi(1)&=\underbrace{T_0^{-1} D_0^{-1} U_0^\dagger}_{M^{-1}_0} \dot\pi(0) \underbrace{U^{}_0 D_0 T^{}_0}_{M^{}_0}\\            
	 &\vdots 
	\end{split}
	\end{equation}
	 
	This numerical strategy for $\dot\pi(t)$ is indeed efficient, but unfortunately its recursive nature quickly becomes unstable.
	The reason is that as we construct $\dot\pi(t)$ for increasing values of $t$, we multiply different decompositions whose diagonal parts have different size orderings.  
	For example, the decomposition of $\mathcal{N}$ has diagonal elements that are ordered from smallest to largest (since it represents an inverse of a matrix, same as $M_t^{-1}$), while the decomposition of $M_t$ will have diagonal components that are instead ordered largest to smallest.  
	This mismatch in orderings not only destroys the separation of scales, but also destroys the preservation of the ordering of scales. 
	This latter feature is just as important in constructing products of matrices in a stabilized manner. 
	
	As an example we show how instabilities arise from the recursive scheme~\eqref{eq:forces_recursive} more systematically by considering a minimal $2\times 2$-sized problem.  
    We consider \texttt{SVD} and focus on the (non-interacting) case where all factors $M_t$ are diagonal in the same basis because this is the most stable scenario.
	In fact, we choose all $M_t$ equal
	\begin{align}
		M_t &= U \diag\br{\eto{\delta E_1}, \eto{-\delta E_2}} U^\dagger\,,
	\end{align}
	without specifying the unitary matrix $U$.
	Any numerical instabilities present in this limit will also appear (likely more severely) in general.
	The only relevant ingredient is a scale separation of energies $E_{1,2} > 0$ so that
	\begin{align}
		\eto{-\delta E_2} \lesssim 1 \lesssim\eto{\delta E_1}\quad;\quad\eto{-\beta E_2} \ll 1 \ll\eto{\beta E_1}\,.
	\end{align}
	For matrices of higher dimension these terms can be thought of as block matrices that group together all positive and all negative energies.
	
	With these assumptions, we can directly calculate
	\begin{equation}
	\begin{split}
				\mathcal{N} &= \left(\mathbb{1}+ U \diag\br{\eto{-\beta E_1}, \eto{\beta E_2}} U^\dagger \right)^{-1}\\
					&= U \diag\br{\frac{1}{1+\eto{-\beta E_1}},\frac{1}{1+\eto{\beta E_2}}} U^\dagger\\
					&= U \diag\br{1, \eto{-\beta E_2}} U^\dagger + \ordnung{\eto{-\beta E_1}} + \ordnung{\eto{-\beta E_2}}\,.
	\end{split}
	\end{equation}
	The crucial step now is to realise that on a computer with finite precision (floating-point) arithmetic the product
	\begin{align}
		U^\dagger U &= \matr{cc}{1 & \epsilon\\ \epsilon & 1}
	\end{align}
	is not exactly the identity matrix.
	Instead, we get deviations dictated by the machine precision $\epsilon$.
	
	The rest of this derivation proceeds by induction.
	Neglecting terms of order $\ordnung{\eto{-\beta E_{1,2}}}$, we write
	\begin{equation}
	\begin{split}
		U \matr{cc}{a_{0} & b_{0}\\ c_{0} & d_{0}} U^\dagger &\coloneqq	\dot \pi(0) = \mathcal{N}\\
		&\approx U \diag\br{1, \eto{-\beta E_2}} U^\dagger\\
		U \matr{cc}{a_{t} & b_{t}\\ c_{t} & d_{t}} U^\dagger &\coloneqq \dot\pi(t) = M^{-1}_{t-1} \dot\pi(t-1) M^{}_{t-1}\\
		&=  M^{-1}_{t-1} U \matr{cc}{a_{t-1} & b_{t-1}\\ c_{t-1} & d_{t-1}} U^\dagger M^{}_{t-1}\,.
	\end{split}
	\end{equation}
	The matrix coefficients $a_t,b_t,c_t,d_t$ are constructed recursively
	\begin{equation}
	\begin{split}
		\matr{cc}{a_{t} & b_{t}\\ c_{t} & d_{t}} &= \diag\br{\eto{-\delta E_1}, \eto{\delta E_2}} \brr{U^\dagger U} \matr{cc}{a_{t-1} & b_{t-1}\\ c_{t-1} & d_{t-1}} \brr{U^\dagger U} \diag\br{\eto{\delta E_1}, \eto{-\delta E_2}}\\
		&= \diag\br{\eto{-\delta E_1}, \eto{\delta E_2}} \brr{\matr{cc}{1 & \epsilon\\ \epsilon & 1} \matr{cc}{a_{t-1} & b_{t-1}\\ c_{t-1} & d_{t-1}} \matr{cc}{1 & \epsilon\\ \epsilon & 1}} \diag\br{\eto{\delta E_1}, \eto{-\delta E_2}}\\
		&= \diag\br{\eto{-\delta E_1}, \eto{\delta E_2}} \left(
		\begin{array}{cc}
			a_{t-1}+\epsilon  \left(b_{t-1}+c_{t-1}\right) & b_{t-1} + \epsilon  \left(a_{t-1}+d_{t-1}\right) \\
			c_{t-1} + \epsilon  \left(a_{t-1}+d_{t-1}\right) & d_{t-1} + \epsilon  \left(b_{t-1}+c_{t-1}\right) \\
		\end{array}
		\right) \diag\br{\eto{\delta E_1}, \eto{-\delta E_2}} + \ordnung{\epsilon^2}\\
		&= \left(
		\begin{array}{cc}
			a_{t-1}+\epsilon  \left(b_{t-1}+c_{t-1}\right) & \eto{-\delta(E_1+E_2)} \br{b_{t-1} + \epsilon  \left(a_{t-1}+d_{t-1}\right)} \\
			\eto{\delta(E_1+E_2)}\br{c_{t-1} + \epsilon  \left(a_{t-1}+d_{t-1}\right)} & d_{t-1} + \epsilon  \left(b_{t-1}+c_{t-1}\right) \\
		\end{array}
		\right) + \ordnung{\epsilon^2}\,.
	\end{split}
	\end{equation}
	In particular, $c_t$ grows exponentially with $t$ while in this example the exact value is $c_t=0$ at all times.
	Simplifying the diagonal terms to their exact values $d_t \ll a_t = 1\ \forall\  t$, we obtain
	\begin{equation}
	\begin{split}
		c_t &\approx \eto{\delta(E_1+E_2)}\br{c_{t-1} + \epsilon}\\
			&= \epsilon \sum_{t'=0}^t \eto{\delta(E_1+E_2) t'}\\
		\Rightarrow |c_t| &> |\epsilon|\, \eto{\delta(E_1+E_2) t}\,.
	\end{split}
	\end{equation}
	Thus, the largest element-wise error is at least
	\begin{align}
		\abs*{c_{N_t-1}} &> |\epsilon|\, \eto{\beta(E_1+E_2)}
	\end{align}
	leading to the determinant
	\begin{align}
		\det \mathcal{N} &= \eto{-\beta E_2}\br{1 + \epsilon c_{N_t-1} + \ordnung{\epsilon^2}}\,.
	\end{align}
	In double-precision arithmetic this implies a total breakdown of the algorithm for $\abs*{\epsilon c_{N_t-1}} \approx 1$ at around $\beta(E_1+E_2) \approx 70$.
	This is consistent with our observations in practice (using $E_1\approx E_2\approx K=3$) where we find that for $\beta \gtrsim 12$ we need a stabilised routine for the calculation of the forces.
	
	This instability is a result of the mixing of size orderings of the diagonal matrices in $M_t$ and $M^{-1}_t$ that occur as we build our recursion in~\eqref{eqn:recursive}.
	If we instead preserve these orderings, for example by \emph{omitting} $M_t^{-1}$ in our recursion and simply calculating $\prod_{t'=0}^t M_{t'}$, stability would be greatly enhanced.  
	A similar calculation as above would give in this case
	\begin{align}
		\abs*{c_{N_t-1}} &\approx |\epsilon|\, \eto{\beta E_1}\\
		&\ll \eto{\beta E_1}\\
		&\approx a_{N_t-1}\,.
	\end{align}
	This means that the off-diagonal terms $c_t$ might still be large on an absolute scale, but they are negligible compared to the diagonal terms $a_t$.
	In consequence, the obtained eigen-spectrum is also correct up to unavoidable $\ordnung{\epsilon}$ effects.
	
	\subsection{Stable calculation of $\partial_{\phi_{x,t}}\log\det(M)$}
	\label{sec:stable-grad}
	Therefore, to obtain a stable form for the forces, we must use expressions that do not mix orderings of scales.  
	We thus generalize~\eqref{eqn:N},
	\begin{equation}
	\mathcal{N}(t)\equiv(\mathbb{1}+\underbrace{M^{-1}_{t-1}\ldots M^{-1}_0 M^{-1}_{N_t-1}\ldots M^{-1}_t}_{N_t\ {\rm terms}})^{-1}\quad\quad;\quad\quad \mathcal{N}(0)\equiv \mathcal{N}\,.
	\end{equation}
	The calculation of $\mathcal{N}(t)$ via decompositions is the same as before, but here the ordering of the products of $M_t$ in $\mathcal{N}(t)$ differs by a simple cyclic permutation to the product that occurs in $\mathcal{N}(t-1)$, and so on.
	It is straightforward to show that 
	\begin{equation}
	\dot\pi(t)=\mathcal{N}(t)\ .
	\end{equation}
	Though the calculation of $\mathcal{N}(t)$ is very stable with our decompositions, we no longer have the efficiency of recursion.   
	
	Having to redo repeated decompositions for each permutation of products of $M_t$ occurring in $\mathcal{N}(t)$ is time consuming.
	Fortunately, we can save considerably on time if we instead iteratively precompute decompositions for the following \textit{prefix} and \textit{suffix} terms,
	\begin{equation}\label{eqn:prefix suffix}
	\begin{split}
	\Pi(t)&\equiv M^{-1}_t M^{-1}_{t-1}\ldots M^{-1}_0=U_\pi(t) D_\pi(t) T_\pi(t)\quad\quad;\quad\quad\Pi(-1)\equiv \mathbb{1}\\
	\Sigma(t)&\equiv M^{-1}_{N_t-1} M^{-1}_{N_t-2}\ldots M^{-1}_t=U_\sigma(t) D_\sigma(t) T_\sigma(t)\ .
	\end{split}
	\end{equation}
	We then have
	\begin{equation}\label{eqn:N with prefix suffix}
	\mathcal{N}(t)=\left(\mathbb{1}+\Pi(t-1) \Sigma(t)\right)^{-1}\ ,
	\end{equation}
	which can be computed in a similar manner as in~\eqref{eqn:decomp N}.
	We stress that the construction of $\Pi(t)$ and $\Sigma(t)$ in~\eqref{eqn:prefix suffix}, and their subsequent application in~\eqref{eqn:N with prefix suffix}, always preserve the order of scales, and thus is numerically very stable.  
	Pseudocode implementations of the prefix and suffix construction \eqref{eqn:prefix suffix} and the subsequent computation of force terms \eqref{eqn:N with prefix suffix} are given in algorithm~\ref{alg:prefix_suffix} and algorithm~\ref{alg:greens_functions}, respectively.
	\begin{algorithm*}
		\caption{\texttt{PrefixSuffix} --- Stable computation of the prefix and suffix terms \eqref{eqn:prefix suffix}.}\label{alg:prefix_suffix}
		\Params{number of time slices $N_t$}
		\Input{
			$\{U_i,\,D_i,\,T_i\}_{i=0}^{N_t-1}$: UDT factors of the matrices $\{M_i^{-1}\}_{i=0}^{N_t-1}$ ($U_i$ unitary, $D_i$ diagonal, $T_i$ upper triangular) \\
			$\mathbb{1}$: Identity matrix
		}
		\Output{
			$U_\pi, D_\pi, T_\pi, U_\sigma, D_\sigma, T_\sigma$: Prefix ($\pi$) and Suffix ($\sigma$) UDT factors for $t=0,\dots,N_t-1$.
		}
		\BlankLine
		\tcp{Initialize first and last element of prefix and suffix, respectively.}
		$(U_\pi[0], D_\pi[0], T_\pi[0]) \gets (U_0, D_0, T_0)$\;
		$(U_\sigma[N_t-1], D_\sigma[N_t-1], T_\sigma[N_t-1]) \gets (U_{N_t-1}, D_{N_t-1}, T_{N_t-1})$\;

		\For{$t\gets 1$ \KwTo $N_t-1$}{
			$(U_\pi[t], D_\pi[t], T_\pi[t]) \gets \FCombineUDT(U_{t}, D_{t}, T_{t}, U_\pi[t-1], D_\pi[t-1], T_\pi[t-1])$\;
			$(U_\sigma[N_t-1-t], D_\sigma[N_t-1-t], T_\sigma[N_t-1-t]) \gets \FCombineUDT(U_\sigma[N_t-t], D_\sigma[N_t-t], T_\sigma[N_t-t], U_{N_t-1-t}, D_{N_t-1-t}, T_{N_t-1-t})$\;
		}
		\KwReturn $U_\pi, D_\pi, T_\pi, U_\sigma, D_\sigma, T_\sigma$\;
	\end{algorithm*}
	\begin{algorithm*}
		\caption{
			\texttt{GreensFunctions} --- Stable computation of the equal time Green's functions \eqref{eqn:eq_time_greens} and unequal time Green's functions \eqref{eqn:uneq_time_greens}.
		}\label{alg:greens_functions}
		\Params{number of time slices $N_t$}
		\Input{
			$U_\pi, D_\pi, T_\pi, U_\sigma, D_\sigma, T_\sigma$: Prefix ($\pi$) and Suffix ($\sigma$) UDT factors for $t=0,\dots,N_t-1$ \\
			\ $\mathbb{1}$: Identity matrix
		}
		\Output{
			$\mathcal{N}_t,G_{tt}, G_{t0}, G_{0t}$: force terms, equal-time ($tt$) and unequal-time ($t0$, $0t$) Green's functions.
		}
		\BlankLine
		\tcp{Equal time Green's functions.}
		$(Q,R) \gets \FQR\bigl( \mathbb{1} + D_\sigma[0] T_\sigma[0] U_\sigma[0] \bigr)$\;
		$\mathcal{N}_{t}[0] \gets U_\sigma[0] R^{-1} (U_\sigma[0] Q)^\dagger$\;
		$G_{tt}[0] \gets \mathbb{1} - \mathcal{N}_t[0]$\;
		\For{$t\gets 1$ \KwTo $N_t-1$}{
			$\tilde{U}, \tilde{D}, \tilde{T} \gets \FCombineUDT\bigl( U_\pi[t-1], D_\pi[t-1], T_\pi[t-1], U_\sigma[t], D_\sigma[t], T_\sigma[t] \bigr)$\;
			$(Q,R) \gets \FQR\bigl( \mathbb{1} + \tilde{D} \tilde{T} \tilde{U} \bigr)$\;
			$\mathcal{N}_{t}[t] \gets \tilde{U} R^{-1} (\tilde{U} Q)^\dagger$\;
			$G_{tt}[t] \gets \mathbb{1} - \mathcal{N}_t[t]$\;
		}
		\BlankLine
		\BlankLine
		\tcp{Unequal time Green's functions $G_{t0}$}
		$G_{t0}[0] \gets G_{tt}[0]$\;
		\For{$t\gets 1$ \KwTo $N_t-1$}{
			$(Q,R) \gets \FQR\bigl( U_\pi^\dagger[t-1] T_\sigma^{-1}[t] D_\sigma^{-1}[t] + D_\pi[t-1] T_\pi[t-1] U_\sigma[t] \bigr)$\;
			$G_{t0}[t] \gets U_\sigma[t] R^{-1} (U_\pi[t-1] Q)^\dagger$\;
		}
		\BlankLine
		\BlankLine
		\tcp{Unequal time Green's functions $G_{0t}$}
		$G_{0t}[0] \gets G_{tt}[0]$\;
		\For{$t\gets 1$ \KwTo $N_t-1$}{
			$(Q,R) \gets \FQR\bigl( U_\sigma^\dagger[t] T_\pi^{-1}[t-1] D_\pi^{-1}[t-1] + D_\sigma[t] T_\sigma[t] U_\pi[t-1] \bigr)$\;
			$G_{0t}[t] \gets - U_\pi[t-1] R^{-1} (U_\sigma[t] Q)^\dagger$\;
		}
				
		\KwReturn $\mathcal{N}_t, G_{tt}, G_{t0}, G_{0t}$\;
	\end{algorithm*}
		
	The force term calculation shares the same \ordnung{N_t N_x^3} scaling established above.
	The overhead from computing the decompositions, the pre- and suffixes $\Pi(t)$ and $\Sigma(t)$, and the stable inversions~\eqref{eqn:N with prefix suffix} is just a constant factor smaller than 10.
	There is an increase in memory demands from storing all the $\Pi(t)$ and $\Sigma(t)$ to a total of $\ordnung{N_t N_x^2}$ which is very rarely a bottleneck.%
	\footnote{A quick example calculation reveals that the total memory required for a simulation with $N_t=256$ and $N_x=\num{2000}$ using double precision amounts to ca.\ $2N_tN_x^2\cdot \SI{8}{\byte}\approx \SI{15}{\giga\byte}$ and thus fits on most modern laptops with $\SI{16}{\giga\byte}$ of RAM or more.
	We have also estimated that for $N_x=\num{2000}$ our current implementation with cubic runtime $\ordnung{N_x^3}$ would lead to about $\num{e6}$ CPU-hours required for a single simulation with just $\num{100}$ configurations (we needed ca.\ 1 CPU-hour for $N_x=20$).
	Thus, in any reasonable scenario runtime becomes a bottleneck long before memory.}
	We also remark that storing all the $UDT$-decompositions of the time slices $M_t$ is exactly as heavy on memory as directly storing the prefix and suffix terms~\eqref{eqn:prefix suffix}. 
		
	\begin{figure}
	\centering
	\includegraphics[width=.8\textwidth]{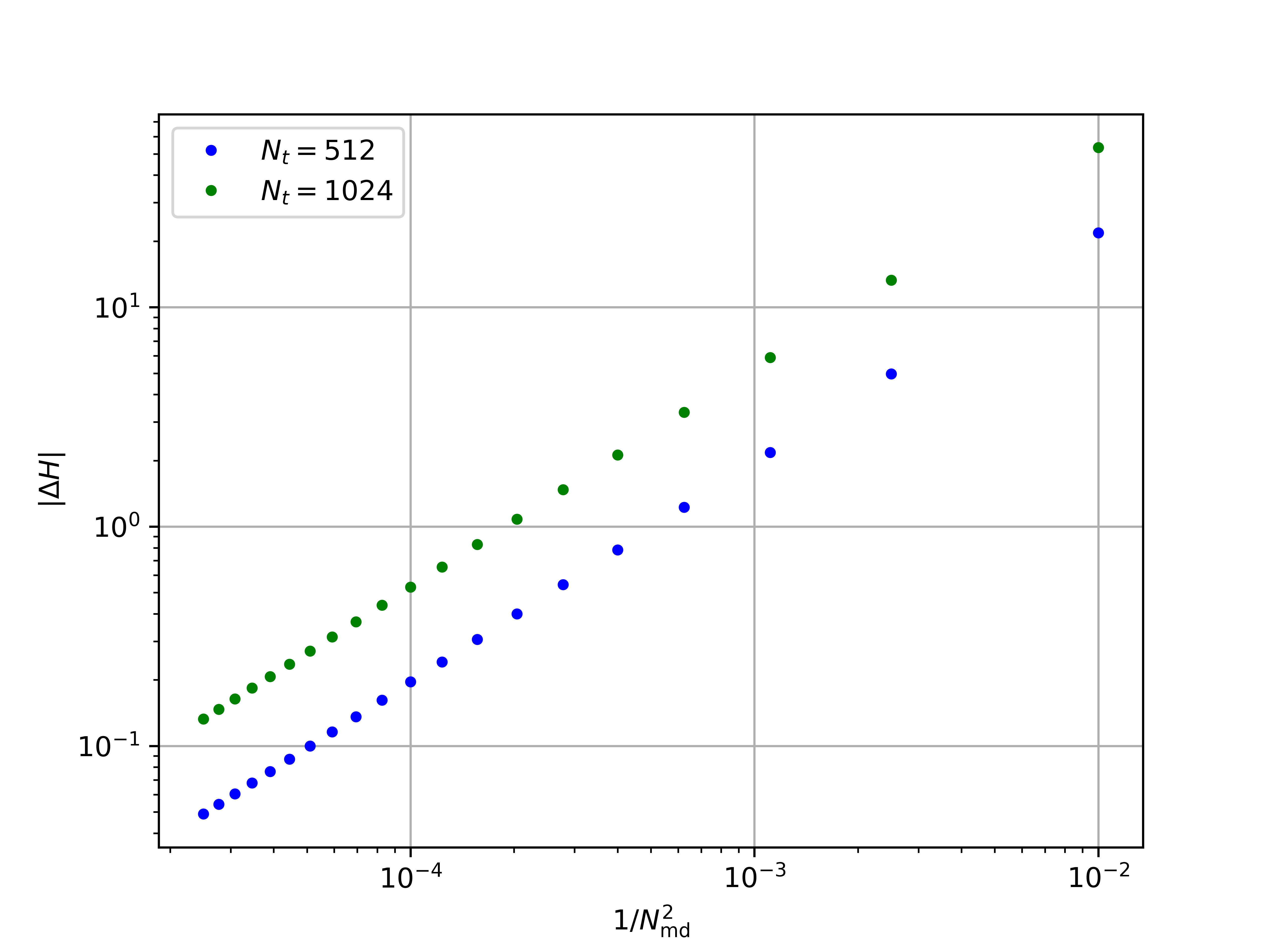}
	\caption{Convergence of leapfrog integrator used in HMC simulations of the Perylene molecule (see e.g.~\cite{Rodekamp:2024ixu}) with onsite coupling $U=2$ and inverse temperature $\beta=100$ (corresponding to room temperature). The change in energy $\Delta H$ of a single HMC trajectory is shown as a function of the inverse number of molecular dynamics steps $N_{\text{md}}$. Calculations were done with two different numbers of time slices, $N_t=512,1024$. The auxiliary field was initially sampled from a Gaussian distribution, $\phi_{x,t}\sim\mathcal{N}(0,\sqrt{\delta U})$.  The expected convergence is $\Delta H\propto N_{\text{md}}^{-2}$.}
	\label{fig:leapfrog convergence}
	\end{figure}
	
\begin{figure}
	\centering
	\includegraphics[width=.8\textwidth]{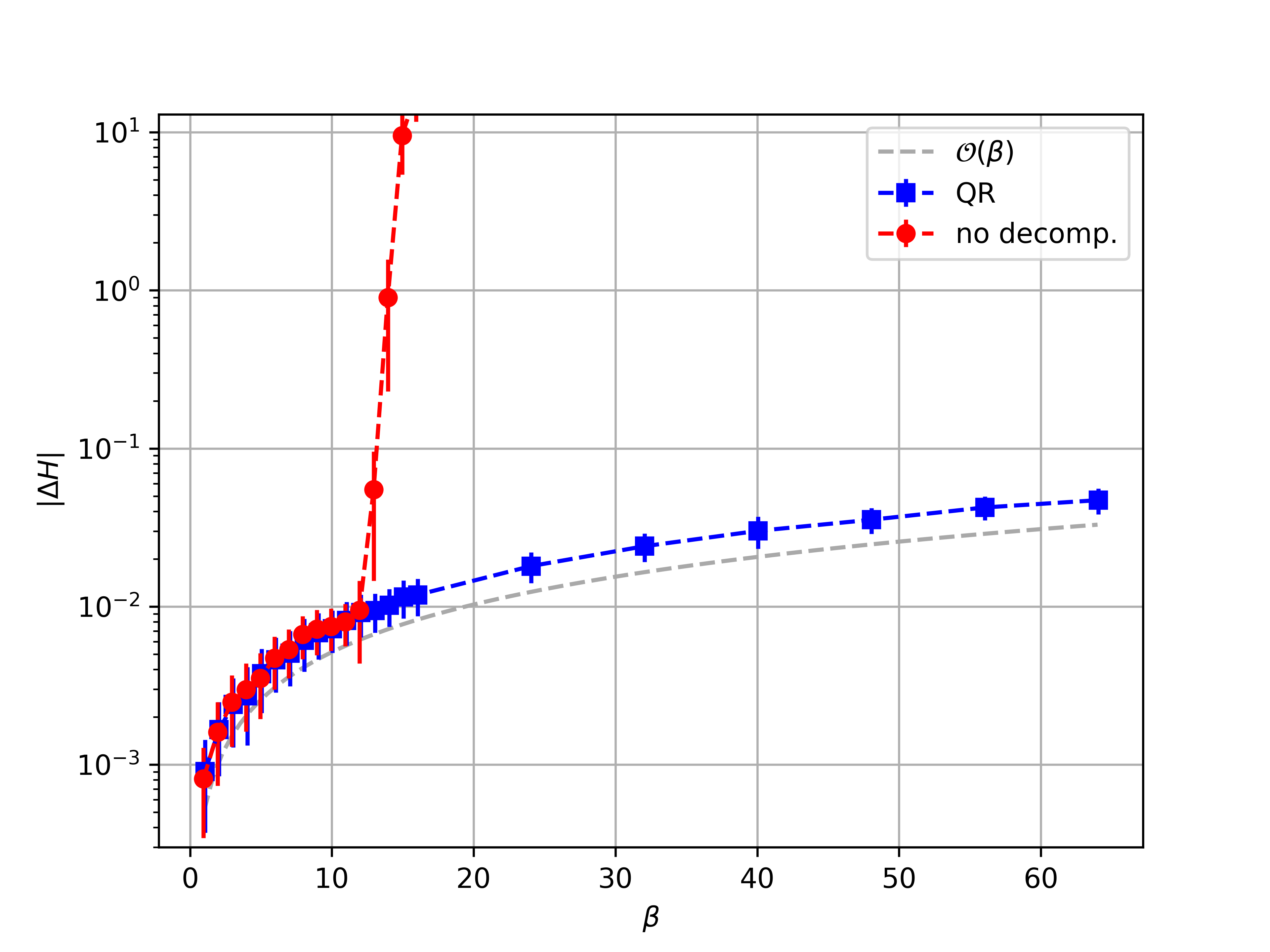}
	\caption{The error incurred during integration of equations of motion during a single HMC trajectory, using our decomposition (``\texttt{QR}'') and na\"ive matrix multiplications (``no decomp.'').  For each data point the trajectory length was the same and the number of leapfrog integration steps was constant with $N_{\text{md}}=50$.  The calculation was repeated 400 times with different random seeds and the median was calculated.  Error bars represent the median absolute deviation.  $N_t$ was chosen such that $\beta/N_t=1/4$. The auxiliary field was initially sampled from a Gaussian distribution, $\phi_{x,t}\sim\mathcal{N}(0,\sqrt{\delta U})$. The expected scaling in the error is $\propto \beta$, and this is shown as the dashed gray line with an arbitrary coefficient.}
	\label{fig:dH convergence}
	\end{figure}

	In fig.~\ref{fig:leapfrog convergence} we show the convergence behavior of the molecular dynamics (MD) `leapfrog' integrator used in HMC simulations for the Perylene system~\cite{Rodekamp:2024ixu} using onsite coupling $U=2$ and inverse temperature $\beta=100$.
	This would correspond to room temperature for a hopping parameter $\kappa=2.7$ eV. 
	The error of the integrator $\Delta H$ should scale with the squared inverse number of MD steps $\mathcal{O}(N_{\text{md}}^{-2})$.
	As can be seen from the figure, this is indeed the case.
	Without decompositions the largest stable $\beta$ attainable that demonstrated appropriate convergence was $\beta\approx 12$~\cite{Rodekamp:2024ixu}.  
	Figure~\ref{fig:dH convergence} shows the error $|\Delta H|$ accumulated during a single HMC trajectory for a 4-site honeycomb system where the trajectory length and number of MD steps were held constant, but the value of $N_t$ varied such that $\beta/N_t=1/4$.  The expected error scales with $\beta$ and this is shown as the dashed gray line in the figure with an arbitrary constant.  It is clear that the decompositions stabilize the trajectory evolution for much larger values of $\beta$, and that na\"ive matrix multiplications result in an error $|\Delta H|\sim10$ at $\beta\approx 15$.
	
	\subsection{Stable calculation of the full propagator $M^{-1}\equiv G$}
	\label{sec:stable-prop}
	We now demonstrate how the terms we have presented in the previous sections can be used to obtain various elements of the Green's function $G$, which is equivalent to the inverse of the fermion matrix~\eqref{eqn:fermion matrix}.
	To do this, we make a further generalization of~\eqref{eqn:prefix suffix} that can be recursively decomposed,
	\begin{equation}\label{eqn:Mtt}
	\mathcal{M}(t',t)=
	\begin{cases}
	M^{-1}_{t'} M^{-1}_{t'-1}\ldots M^{-1}_{t+1} M^{-1}_{t}\quad &\forall\ t'\ge t\\
	M^{-1}_{(t'+N_t)\%N_t} M^{-1}_{(t'+N_t-1)\%N_t}\ldots M^{-1}_{(t+1)\%N_t} M^{-1}_{t}\quad &\forall\ t'< t
	\end{cases}\ .
	\end{equation}
	The expressions in~\eqref{eqn:prefix suffix} are related to~\eqref{eqn:Mtt} by $\Pi(t)=\mathcal{M}(t,0)$ and $\Sigma(t)=\mathcal{M}(N_t-1,t)=\mathcal{M}(-1,t)$.
	These terms allow for a very compact expression of the propagator,
	\begin{equation}\label{eqn:general_greens_function}
	G(t_f,t_i)=\mathcal{B}_{t_f,t_i}\times
	\begin{cases}
	[\mathbb{1}+\mathcal{M}^{-1}(t_i-1,t_i)]^{-1} & \forall\ t_f = t_i\\
	\left[\mathcal{M}^{-1}(t_i-1,t_f)+\mathcal{M}(t_f-1,t_i)\right]^{-1} & \forall\ t_f \ne t_i
	\end{cases}\ ,
	\end{equation}
	where
	\begin{equation}
	\mathcal{B}_{t_f,t_i}=
	\begin{cases}
	+1 & \forall\ t_f\ge t_i\\
	-1 & \forall\ t_f<t_i
	\end{cases}\ .
	\end{equation}
	Note that $G(t_f,t_i)$ represents an $N_x\times N_x$ matrix. 
	The total runtime to obtain the entire Green's function scales as $\ordnung{N_x^3 N_t^2}$ which is a necessary consequence of $N_t^2$ different $t_i,t_f$ combinations of spatially dense matrices.
	
	If we now concentrate on the diagonal $t_f=t_i=t$ elements, the runtime drops back to $\ordnung{N_x^3 N_t}$.
	In that case we have
	\begin{equation}
	\label{eqn:eq_time_greens}
	G(t,t)	=[\mathbb{1}+\mathcal{M}^{-1}(t-1,t)]^{-1} 
			=[\mathbb{1}+\Sigma^{-1}(t) \Pi^{-1}(t-1)]^{-1} 
			=\mathbb{1} - [\mathbb{1}+\Pi(t-1) \Sigma(t)]^{-1} 
			= \mathbb{1} - \mathcal{N}(t)\ ,
	\end{equation}
	which is easily obtainable from the force terms of~\eqref{eqn:N with prefix suffix}.
	Thus as we calculate the force terms needed for our HMC evolution, we automatically obtain the diagonal (in time) elements of the fermion propagator.
	Further inspection shows that during an HMC evolution we calculate terms that allow us to construct the first column and first row of the fermion propagator, $G_{t,0}$ and $G_{0,t}$ respectively,
	\begin{align}
		\label{eqn:uneq_time_greens}
		\begin{split}
			G(t,0)=&\left[\mathcal{M}^{-1}(-1,t)+\mathcal{M}(t-1,0)\right]^{-1}=\left[\Sigma^{-1}(t)+\Pi(t-1)\right]^{-1}\,,\\
			G(0,t)=&-\left[\mathcal{M}^{-1}(t-1,0)+\mathcal{M}(-1,t)\right]^{-1}=-\left[\Pi^{-1}(t-1)+\Sigma(t)\right]^{-1}\,.
		\end{split}
	\end{align}
	Using~\eqref{eqn:prefix suffix} we can express these terms as
	\begin{equation}
	\begin{split}
	G(t,0)=&\left[T^{-1}_\sigma(t) D^{-1}_\sigma(t) U^\dagger_\sigma(t)+U_\pi(t-1) D_\pi(t-1) T_\pi(t-1)\right]^{-1}\,,\\
	G(0,t)=&-\left[T^{-1}_\pi(t-1) D^{-1}_\pi(t-1) U^\dagger_\pi(t-1)+U_\sigma(t) D_\sigma(t) T_\sigma(t)\right]^{-1}\,.
	\end{split}
	\end{equation}
	These expressions need to be decomposed one remaining time.
	For example, for $G(t,0)$ we perform
	\begin{multline}
	G(t,0)=\left[T^{-1}_\sigma(t) D^{-1}_\sigma(t) U^\dagger_\sigma(t)+U_\pi(t-1) D_\pi(t-1) T_\pi(t-1)\right]^{-1}\\
	=\left[U_\pi(t-1) \underbrace{\left(U^\dagger_\pi(t-1) T^{-1}_\sigma(t) D^{-1}_\sigma(t) +D_\pi(t-1) T_\pi(t-1)U_\sigma(t)\right)}_{udt}U^\dagger_\sigma(t)\right]^{-1}\\
	=U_\sigma(t)t^{-1}d^{-1}(U_\pi(t-1)u)^\dagger \ .\label{eq:sym_decomp_Gt0}
	\end{multline}
	A similar decomposition is performed for $G(0,t)$.
	Pseudocode for the construction of $G(t,t)$, $G(t,0)$, and $G(0,t)$ is given in algorithm~\ref{alg:greens_functions}.
	
	The form of this \emph{symmetric} decomposition, where the unitary matrices $U_\pi$ and $U^\dagger_\sigma$ are factored out,  is motivated by the fact that in the non-interacting limit and using \texttt{SVD}, we have that $U_{\sigma,\pi}=T^{-1}_{\sigma,\pi}\ \forall\ t$, giving
	\begin{displaymath}
	U^\dagger_\pi(t-1) T^{-1}_\sigma(t) D^{-1}_\sigma(t) +D_\pi(t-1) T_\pi(t-1)U_\sigma(t)=D^{-1}_\sigma(t) +D_\pi(t-1) \quad\quad (\mbox{non-interacting})\ .
	\end{displaymath}
	This expression is purely diagonal and thus maximally stable under decompositions.
	In the case of \texttt{QR} the stability of this decomposition is less obvious.
	The re-ordered terms $U^\dagger T^{-1} D^{-1}$ and $D T U$ feature a compelling symmetry and are again reminiscent of the first step in a QR algorithm (see discussion following eq.~\eqref{eqn:decomp3}).
	While this in general does not  guarantee stability, we have verified that we obtain high accuracy in multiple numerical experiments.
	We have further tested all simple alternative re-orderings of the $UDT$ factors and found this decomposition superior or at least equivalent in all our tests.
	To demonstrate the stability of calculating the Green's function with these decompositions, we show examples in fig.~\ref{fig:Gxy_comp} of the propagator obtained for the 8-site honeycomb system with periodic boundary conditions in the non-interacting limit using our recursive decompositions in~\eqref{eq:sym_decomp_Gt0} and without using any stabilization method, i.e.\@ direct matrix multiplications.
	\begin{figure}
	\includegraphics[width=\textwidth]{Gxy_comp.png}
	\caption{Comparison of the spatial propagator calculated using eq.~\eqref{eq:sym_decomp_Gt0} (orange dots) and using no stabilization (blue squares) with exact results (solid line) for the 8-site honeycomb lattice with periodic boundary conditions.  The three distinct propagators (see text) are separated by rows, and the different $\beta$ are separated by columns.  In the $\beta=100$ column only the results obtained via decompositions are shown since calculations without stabilizations numerically failed.  There is an obvious benefit to using recursive decompositions for $\beta \gtrsim 12$.
	\label{fig:Gxy_comp}}
	\end{figure}
	Due to the symmetry of this system, there are only three distinct spatial propagators corresponding to equal site propagation $G_{x,x}(t,0)$, nearest-neighbor $G_{\langle x,y\rangle}(t,0)$, and next-to-nearest-neighbor $G_{\langle\langle x,y\rangle\rangle}(t,0)$ propagation.  These different propagators are represented by the different rows of fig.~\ref{fig:Gxy_comp}.
	In this non-interacting limit we can obtain the exact spatial propagators, which allows us to test the correctness of our stabilizations with different $\beta$s which are demarcated by columns in fig.~\ref{fig:Gxy_comp}.
	As can be seen from the figure, our decompositions provide stablization for all $\beta$s shown, whereas the naïve calculation of the propagator starts to fail already at $\beta=12$, in agreement with our findings in fig.~\ref{fig:dH convergence}.	For the $\beta=100$ case the calculation without stabilization is no longer possible due to excessive precision loss, and so only the propagator obtained via stabilization is shown.
	
\begin{figure}
	\includegraphics[width=\textwidth]{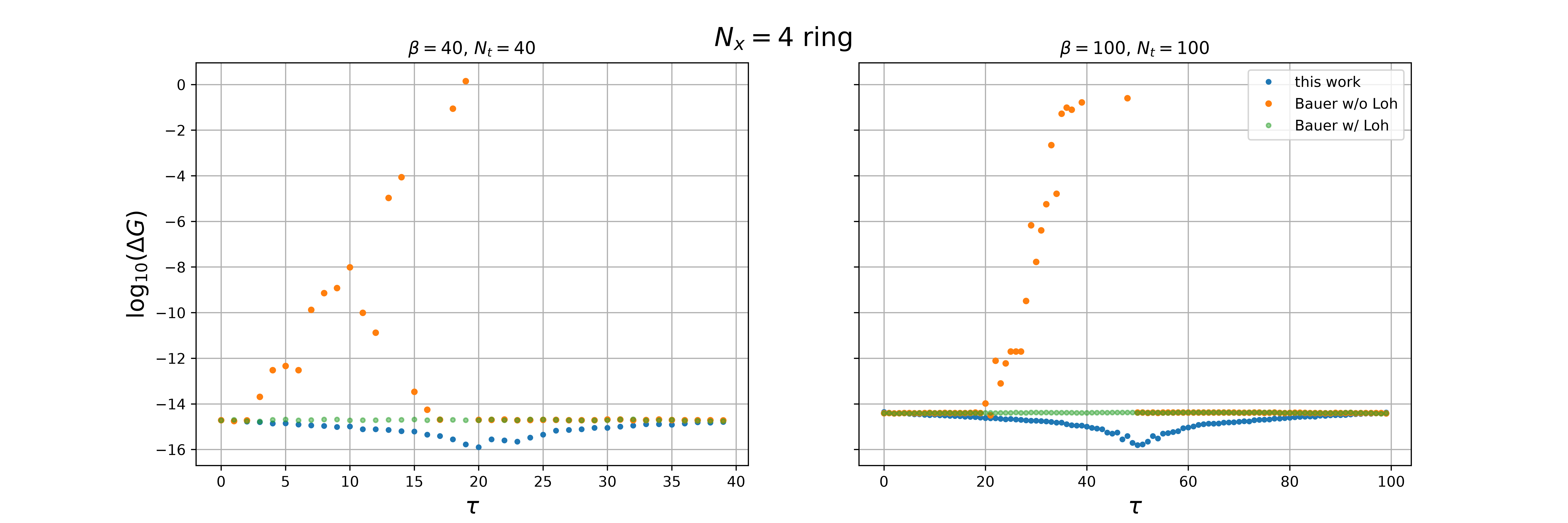}
	\caption{Accuracy of various decompositions obtained from the calculation of the 4-site ring propagator as a function of imaginary time $\tau$.  What is shown is $\Delta G = \log_{10}\operatorname{max}\operatorname{abs}\left(G(\tau)-G_\text{exact}(\tau)\right)$ in the $U=0$ limit, where $G_\text{exact}(\tau)$ is the exact propagator obtained from exact diagonalization (similar to Fig.~6 of~\cite{Bauer:2020stable}).  The \texttt{Bauer w/o Loh} and \texttt{Bauer w/ Loh}  results refer to eq.(28) and eq.(29) of~\cite{Bauer:2020stable}, respectively, and employ \texttt{QR} decompositions with pivoting. \label{fig:Bauer vs Us}}
	\end{figure}

	In Bauer's paper~\cite{Bauer:2020stable} a slightly less symmetric decomposition is applied in the last step to obtain the time-displaced Green's function (see Eq.(28) of this paper).  Bauer finds that for growing $\beta$ this decomposition is also unstable and must be augmented by an additional Loh splitting~\cite{Loh:2005}.\footnote{We note that there is a typo in Eq.(29) of~\cite{Bauer:2020stable}. The third line of this equation should read
	\begin{equation}
	\left[U_LD_{L_p}\left(D_{L_m}X^{}_LX^{-1}_R D^{-1}_{R_p}+D^{-1}_{L_p}U^\dagger_LU^{}_RD_{R_m}\right)D_{R_p}X_R\right]^{-1}\ .
	\end{equation}
	The typo does not affect the results of~\cite{Bauer:2020stable}, since the code~\cite{Bauer:code} used to obtain them implements the correct expression.
	}  This in turn incurs an extra decomposition, making the algorithm slightly less efficient.  We find that, with our symmetric formulation our algorithm remains stable for very high $\beta$, performing as well in accuracy as with  Bauer's Loh-augmented decomposition.  Figure~\ref{fig:Bauer vs Us} demonstrates this accuracy by computing the spatial propagator for the 4-site ring using these different decompositions and comparing them with the exact correlator in the non-interacting limit.  These comparisons were done with both $\beta=40$ and $\beta=100$.  In both cases our decomposition is as accurate as (and in many cases slightly more accurate than) the Loh-method, with the added benefit of being also more efficient.
	
\begin{figure}
	\centering
	\includegraphics[width=.8\textwidth]{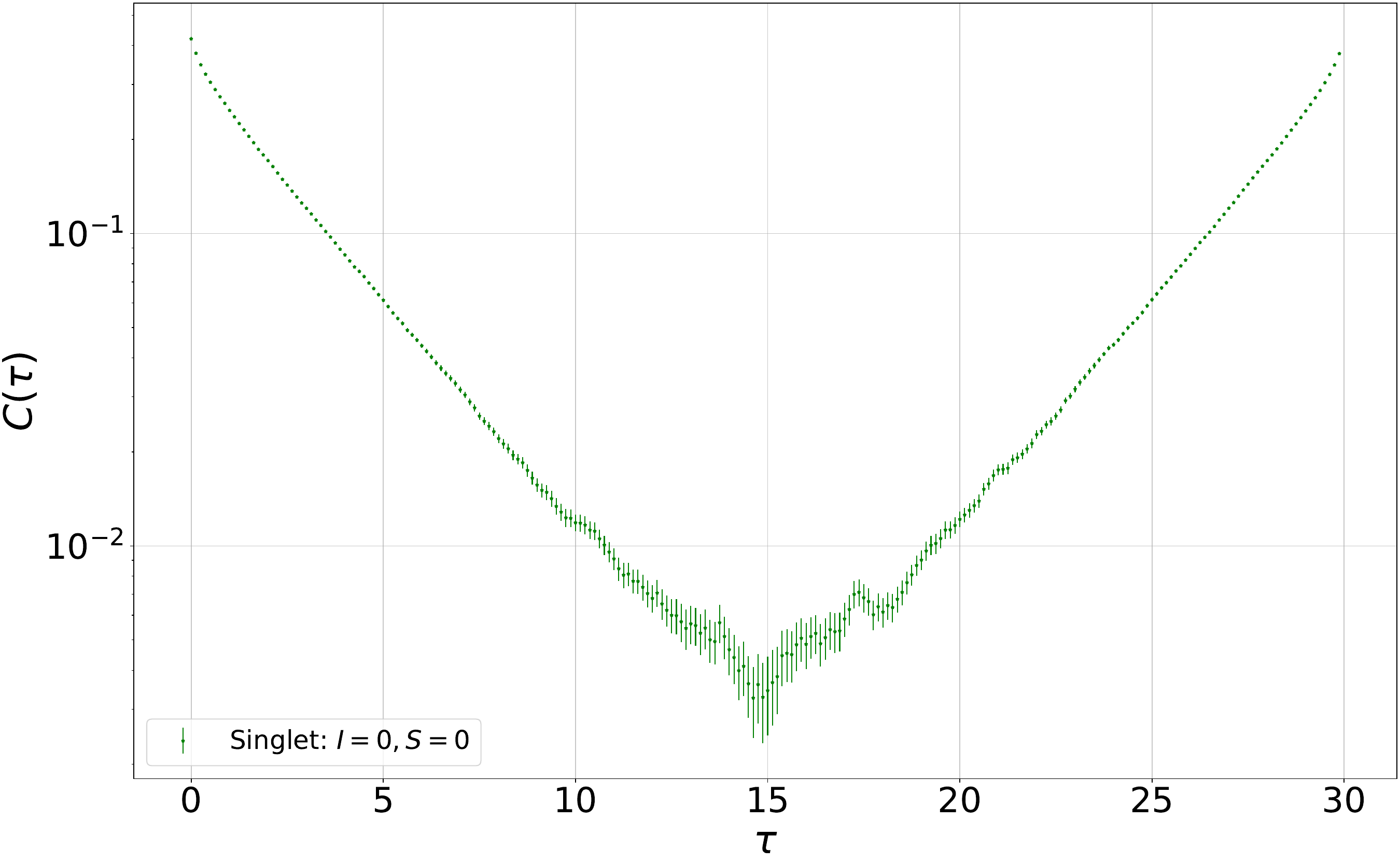}
	\caption{Two-body correlator $C(\tau)$ for a spin-singlet $(S=0)$ bright exciton, corresponding to a particle-hole excitation, in a $(10,2)$ chiral carbon nanotube ($N_x = 248$) at inverse temperature $\beta=30$. The channel shown has total pseudo-spin $I=0$. The data are obtained from HMC simulations using our decomposition method, which enables stable measurements of the excitonic correlator at large inverse temperature. \label{fig:exciton}}
	\end{figure}

	Having direct access to these components of the Green's function means that we can calculate \emph{any} \emph{multi-}body two-point Green's function with initial time $t_i=0$ and final time $t_f=t$ arbitrary.  For example, the expectation value of the following general two-point operator,
	\begin{displaymath}
	\langle O^{}(t) O^\dagger(t_i=0)\rangle
	\end{displaymath}
	where $O^{}(t)$ is a product of $N$ fermion creation ${\bf c}^\dagger(t)$ and/or annihilation ${\bf c}^{}(t)$ operators\footnote{Bold symbols denote spatial vectors of operators, e.g.\@ ${\bf c}^{}(t)=(c^{}_0(t),c^{}_1(t),\ldots,c^{}_x(t),\ldots)$ represents a vector of annihilation operators.} at time $t$, upon Wick contraction will result in a linear combination of products of one-body Green's functions of the following types,
	\begin{align}\label{eqn:terms}
	\wick{ \c1 {\bf  c}^{}(t)\c1 {\bf c}^\dagger(0)}&=G(t,0); &
	\wick{ \c1 {\bf c}^{}(0)\c1 {\bf c}^\dagger(t)}&=G(0,t); &
	\wick{ \c1 {\bf c}^{}(0)\c1 {\bf c}^\dagger(0)}&=G(0,0);&
	\wick{ \c1 {\bf c}^{}(t)\c1 {\bf c}^\dagger(t)}&=G(t,t)\ .
	\end{align}
	Here the overhead brackets represent a Wick contraction.
	Note that each of these Green's functions  is implicitly a functional of the auxiliary field $\phi$.  The `disconnected' term $G(t,t)$ has historically been difficult to calculate, and has instead been approximated stochastically (see, e.g.\@ Ref.~\cite{Ostmeyer:2021efs}).
	On the other hand, each of these terms is calculated with our decompositions during an HMC evolution without approximation.  
	By appropriate combinations and products of these terms we can therefore construct the $N$-body two-point Green's function. 
	For example, we show the 2-body two-point time-dependent Green's function, or correlator $C(t)$, representing a spin- and pseudospin-singlet ``bright'' exciton (particle/hole excitation) with zero total momentum (i.e.\@ $\Gamma$-point) coming from a $(10,2)$ chiral carbon nanotube~\cite{Luu:2015gpl} in fig.~\ref{fig:exciton}.
	The contraction resulting in this correlator is given in app.~\ref{sect:contraction} (eq.~\eqref{eqn:contraction}) where we see that all terms of~\eqref{eqn:terms}, and in particular the disconnected terms, are required. 	

	\section{Conclusion}
	
	We have derived an algorithm to calculate the fermion Green's function accurately with high numerical stability in determinant quantum Monte Carlo (DQMC) simulations.
	This algorithm paves the way to low temperature simulations.
	As demonstrated in \cref{fig:leapfrog convergence}, simulations at inverse temperatures of $\beta=100$ corresponding to room temperature in carbon nano systems are now feasible.
	Without a stabilising procedure, on the other hand, graphene structures at $\beta\gtrsim 15$ are practically impossible to simulate (see fig.~\ref{fig:dH convergence}).
	More generally, depending on the band width $\Delta$, the na\"ive implementation using double-precision arithmetic breaks down at around $\beta\Delta \approx 70$.
	At the same time, our algorithm has a runtime complexity of $\ordnung{N_x^3 N_t}$ which is exactly equivalent to that of the fastest possible (unstable) implementation using dense spatial matrices.
	
	In a nutshell, the main concepts used in our algorithm are as follows.
	We first realise that all the fermionic dynamics are encoded in $N_t$ dense spatial $N_x\times N_x$ matrices $M_t$.
	Moreover, each element of the fermionic Green's function can be written in the form~\eqref{eqn:general_greens_function} containing only products of time slices $M_t$, one addition and a final stable inversion~\eqref{eq:sym_decomp_Gt0}.
	All the required cyclic permutations of $\prod_t M_t$ can be obtained efficiently by storing only $\ordnung{N_t}$ pre- and suffix matrices~\eqref{eqn:prefix suffix}.
	The procedure is stabilised by splitting each $M_t=U_tD_tT_t$ using a \texttt{QR} decomposition with unitary $U_t$, diagonal $D_t$ and triangular $T_t$.
	Every matrix subsequently constructed from these building blocks is stored in the same $UDT$ format.
	Careful treatment of the diagonals $D$ ensures a consistent ordering and separation of scales so that precision loss can be avoided.	
	This procedure also allows us to calculate the fermion determinant and its derivatives like the forces required for HMC simulations, almost as a by-product.
	In practice, stabilising every single matrix product is usually not required to retain sufficient numerical precision, and the procedure generalises straightforwardly to stabilising only on every $n$-th product.
	This further reduces the computational cost and allows the algorithm to be combined with sparse methods, which we will explore in future work.
	A collection of stable and efficient algorithms for dealing with the fermion matrix including sparse methods for larger systems is provided in the handbook~\cite{Ostmeyer:2026sea}.
	
	Our algorithm is inspired by that introduced in Ref.~\cite{Bauer:2020stable}, but it comes with several additional features.
	The pre- and suffix calculation allows us to treat all the Green's function elements on the same footing preserving maximal stability without sacrificing runtime.
	This makes additional complicated procedures like the Loh splitting superfluous.
	We also explain the origins of the instabilities of the na\"ive approach theoretically beyond empirical evidence.
	These error sources are fully eliminated by our algorithm.
	
	The last few years have seen multiple substantial algorithmic improvements particularly relevant for HMC simulations of fermionic systems like the Hubbard model.
	The HMC has been tuned for minimal autocorrelation~\cite{Ostmeyer:2024amv} and combined with radial updates~\cite{original_radial_update} guaranteeing exponentially fast thermalisation~\cite{Ostmeyer:2024gnh} and fully ergodic simulations even in the presence of some potential barriers~\cite{Temmen_ergodicity}.
	In addition, the robust analysis of noisy Euclidean correlators has been simplified~\cite{Ostmeyer:2025igc}.
	All of these improvements will become relevant when tackling ambitious projects like realistic simulations of organic molecules away from half filling which we intend to start in the near future.
	First simulations of such a molecule (Perylene) have been undergone successfully in the presence of a sign problem~\cite{Rodekamp:2024ixu}, albeit at relatively high temperatures.
	With this work we have added the remaining cornerstone that allows us to approach room temperature.
	
	\section*{Code and Data}
	Implementations of all the stabilised routines discussed in this work are available publicly within the \texttt{NSL}~\cite{NSL} library.
	Data will be made available upon reasonable request.
	
	\begin{acknowledgments}
		We are grateful to Felipe Attanasio for drawing our attention to Ref.~\cite{Bauer:2020stable}, which served as an important reference for this work. We also thank Dominic Schuh and Janik Kreit for insightful discussions.
		This work was funded by the Deutsche Forschungsgemeinschaft (DFG, German Research Foundation) as part of the
		CRC 1639 NuMeriQS – Project number 511713970; and MKW NRW under the funding code NW21-024-A. We gratefully acknowledge the computing time
		granted by the JARA Vergabegremium and provided on the JARA Partition part of the supercomputer JURECA at
		Forschungszentrum Jülich~\cite{jureca-2021}.
	\end{acknowledgments}

	\FloatBarrier
	
	\bibliography{bibliography}
	
	\appendix

	\section{The HMC algorithm\label{app:HMC}}
	The HMC algorithm treats the distribution involving $S_\text{eff}[\phi]$ (defined in~\eqref{eqn:partition}),
	\begin{displaymath}
	P[\phi]=\frac{e^{-S_\text{eff}[\phi]}}{\int\mathcal{D}[\phi]e^{-S_\text{eff}[\phi]}}
	\end{displaymath}
	as a marginal distribution obtained from integrating over `conjugate momenta $\pi$' of the distribution
	\begin{displaymath}
	P[\pi,\phi]=\frac{e^{-\frac{1}{2}\sum_{x,t}\pi_{x,t}^2-S_\text{eff}[\phi]}}{\int\mathcal{D}[\pi]\mathcal{D}[\phi]e^{-\frac{1}{2}\sum_{x,t}\pi_{x,t}^2-S_\text{eff}[\phi]}}
	\equiv
	\frac{e^{-\mathcal{H}[\pi,\phi]}}{\int\mathcal{D}[\pi]\mathcal{D}[\phi]e^{-\mathcal{H}[\pi,\phi]}}\ ,
	\end{displaymath}
	where $\mathcal{D}[\pi]=\prod_{x,t}d\pi_{x,t}$ and the `artificial Hamiltonian'
	\begin{equation}
	\mathcal{H}[\pi,\phi]=\frac{1}{2}\sum_{x,t}\pi_{x,t}^2+S_\text{eff}[\phi]=\frac{1}{2}\sum_{x,t}\pi_{x,t}^2+S[\phi]-\sum_{f=1,2}\log\det M^{(f)}[\phi]\ .
	\end{equation}
	For each auxiliary field $\phi_{x,t}$ there is a corresponding conjugate momentum $\pi_{x,t}$.  To generate a new state $\phi_{\text{new}}$, one first samples the conjugate momenta from a normal distribution with unit variance, $\pi_{x,t}\sim \mathcal{N}(0,1)$, and then integrates the EoMs of the artificial Hamiltonian,
	\begin{equation}
	\begin{split}
	\dot\phi_{x,t}&=\partial_{\pi_{x,t}}\mathcal{H}[\pi,\phi]=\pi_{x,t}\\
	\dot\pi_{x,t}&=-\partial_{\phi_{x,t}}\mathcal{H}[\pi,\phi]=-\partial_{\phi_{x,t}}S[\phi]+\sum_{f=1,2}\partial_{\phi_{x,t}}\log\det M^{(f)}[\phi]
	\end{split}\ ,
	\end{equation}
	for some prescribed trajectory length\footnote{For most theories the evaluation of  $-\partial_{\phi_{x,t}}S[\phi]$ is straightforward. For example, for the Hubbard action given in~\eqref{eqn:Hubbard action} it is simply $-\frac{\phi_{x,t}}{U\delta}$.}.  Assuming exact integration, one takes the resulting field after integration of the EoMs as the next state in the Markov chain. The process is repeated to generate the ensemble of configurations.
	
	Exact integration, unfortunately, is not possible except in the most trivial cases.  
	Instead one uses numerical integration methods, like `leap-frog' or the more general `Omelyan' integrators~\cite{OMELYAN2003272,Malezic:2026bds,Ostmeyer:2026ckx}, that are symplectic and thus area preserving, as well as reversible which ensures detailed balance.  The resulting integration has an error $\Delta\mathcal{H}$, and this is used in the Metropolis accept/reject step to determine whether to accept the new proposed state $\phi_\text{new}$ with probability
	\begin{displaymath}
	P_\text{acc}=\min\left(1,e^{-\Delta\mathcal{H}[\pi_\text{new},\phi_\text{new}]}\right)\ ,
	\end{displaymath}
	or to copy the original state into the Markov chain.
	
	\section{2-body two-point exciton contraction\label{sect:contraction}}
	The 2-body two-point Green's function, or correlator, representing a spin- and pseudospin-singlet ``bright'' exciton (particle/hole excitation) with back to back momentum $k$ resulting in zero total momentum ($\Gamma$ point), after Wick contraction, is
	\begin{multline}\label{eqn:contraction}
	\langle O^{}_{k,-k}(t>0)O^\dagger_{k,-k}(0)\rangle\equiv C(t)=\\
	\frac{1}{4}\left \langle
G^{p}_{k,-k}(0,0) G^{h}_{k,-k}(t,t) -G^{p}_{k,-k}(0,0) G^{h}_{-k,k}(t,t) -G^{p}_{-k,k}(0,0) G^{h}_{k,-k}(t,t) +G^{p}_{-k,k}(0,0) G^{h}_{-k,k}(t,t) \right.\\
+G^{h}_{k,-k}(0,0) G^{p}_{k.-k}(t,t) -G^{h}_{k,-k}(0,0) G^{p}_{-k,k}(t,t) -G^{h}_{-k,k}(0,0) G^{p}_{k,-k}(t,t) +G^{h}_{-k,k}(0,0) G^{p}_{-k,k}(t,t) \\
+G^{h}_{k,k}(t,0) G^{h}_{-k,-k}(0,t) -G^{h}_{k,-k}(t,0)  G^{h}_{k,-k}(0,t) +G^{h}_{k,-k}(0,0) G^{h}_{k,-k}(t,t) -G^{h}_{-k,k}(0,0) G^{h}_{k,-k}(t,t) \\
-G^{h}_{-k,k}(t,0) G^{h}_{-k,k}(0,t) +G^{h}_{k,k}(0,t)  G^{h}_{-k,-k}(t,0) -G^{h}_{k,-k}(0,0) G^{h}_{-k,k}(t,t) +G^{h}_{-k,k}(0,0) G^{h}_{-k,k}(t,t) \\
+G^{p}_{k,k}(t,0) G^{p}_{-k,-k}(0,t) -G^{p}_{k,-k}(t,0)  G^{p}_{k,-k}(0,t) +G^{p}_{k,-k}(0,0) G^{p}_{k,-k}(t,t) -G^{p}_{-k,k}(0,0) G^{p}_{k,-k}(t,t) \\
-\left.G^{p}_{-k,k}(t,0) G^{p}_{-k,k}(0,t) +G^{p}_{k,k}(0,t)  G^{p}_{-k,-k}(t,0) -G^{p}_{k,-k}(0,0) G^{p}_{-k,k}(t,t) +G^{p}_{-k,k}(0,0) G^{p}_{-k,k}(t,t)\right\rangle\ ,
	\end{multline}
	where the superscript $p$ ($h$) refers to the particle (hole) Green's function and the $\langle\cdots\rangle$ on the RHS denotes that an average over the auxiliary fields $\phi$ is performed.

	\section{Pseudocode}
	\label{app:pseudocode}
	Notation and conventions for the pseudocode given throughout this work are detailed in Tab.~\ref{tab:pseudocode}.
	\begin{table}
	\centering
		\begin{tabular}{|c | c |}
			\hline
			Symbol & Meaning \\ \hline
			$\gets$ & assignment \\
			$AB$ (juxtaposition) & matrix-matrix product\\
			$\Fdiag(A)$ & extract diagonal of $A$ as matrix\\
			$\log(A)$ & natural logarithm\\
			$\det(A)$ & determinant\\
			$A^{-1}$ & matrix inverse\\
			$A^\dagger$ & conjugate transpose\\
			$(Q,R)\gets\FQR(A)$ & QR decomposition $A = QR$, $Q$ unitary, $R$ upper triangular\\ 
			$A[t]$ & $t$-th element of sequence of matrices, $t=0,1,\dots,N_t-1$ \\ \hline
		\end{tabular}
	\caption{Notation and conventions used in the pseudocode.}
	\label{tab:pseudocode}
	\end{table}
	
\end{document}